\documentclass[aps,reprint,prl,superscriptaddress,amsmath,amssymb]{revtex4-1}
\usepackage{graphicx}
\usepackage{amsmath}
\usepackage{bm}

\usepackage{color}

\definecolor{CLBlue}{rgb}{0, .3, .6}

\begin{document}

\title{Decomposing the local arrow of time in interacting systems}

\author{Christopher W. Lynn}
\affiliation{Initiative for the Theoretical Sciences, The Graduate Center, City University of New York, New York, NY 10016, USA}
\affiliation{Joseph Henry Laboratories of Physics and Lewis–Sigler Institute for Integrative Genomics, Princeton University, Princeton, NJ 08544, USA}
\author{Caroline M. Holmes}
\affiliation{Joseph Henry Laboratories of Physics and Lewis–Sigler Institute for Integrative Genomics, Princeton University, Princeton, NJ 08544, USA}
\author{William Bialek}
\affiliation{Initiative for the Theoretical Sciences, The Graduate Center, City University of New York, New York, NY 10016, USA}
\affiliation{Joseph Henry Laboratories of Physics and Lewis–Sigler Institute for Integrative Genomics, Princeton University, Princeton, NJ 08544, USA}
\author{David J. Schwab}
\affiliation{Initiative for the Theoretical Sciences, The Graduate Center, City University of New York, New York, NY 10016, USA}

\date{\today}

\begin{abstract}
We show that the evidence for a local arrow of time, which is equivalent to the entropy production in thermodynamic systems, can be decomposed. In a system with many degrees of freedom, there is a term that arises from the irreversible dynamics of the individual variables, and then a series of non--negative terms contributed by correlations among pairs, triplets, and higher--order combinations of variables. We illustrate this decomposition on simple models of noisy logical computations, and then apply it to the analysis of patterns of neural activity in the retina as it responds to complex dynamic visual scenes. We find that neural activity breaks detailed balance even when the visual inputs do not, and that this irreversibility arises primarily from interactions between pairs of neurons.
\end{abstract}

%\pacs{asdfasdf}
%\keywords{keywords}

\maketitle


A system held in steady state, away from thermal equilibrium, must continuously dissipate heat to the surrounding bath, causing an increase in entropy. Such a system also violates detailed balance, so that the time reversed trajectories must be measurably less probable than the true trajectories; observation of the system trajectory thus provides evidence for the arrow of time. An important result of modern non--equilibrium statistical mechanics is that the rate at which evidence---in the precise, information--theoretic sense---accumulates for the arrow of time is equal to the rate at which entropy is produced in the bath \cite{Seifert-01, Kawai-01}. This idea has been used to search for signatures of irreversibility in experimental data, notably on a wide range of living systems, across scales from single cells \cite{Battle-01,Gnesotto-01,Tan-01} to global brain dynamics \cite{Lynn-09, Perl-01}. 

In almost all the systems where we want to study entropy production and the arrow of time, there are many interacting degrees of freedom. If we try to estimate the entropy, then we know that treating each variable independently results in an over--estimate, and that as we take account of correlations among pairs, triplets, and larger groups of variables we generate a monotonically decreasing hierarchy of bounds \cite{Schneidman-02}. Here we show that the opposite is true of the entropy production, or the evidence for the arrow of time: we can decompose this measure of irreversibility into a series of non--negative terms, corresponding to successive orders of correlation or interaction among the variables in the system. While correlations always decrease the entropy, we find that correlations always increase the evidence for irreversibility. This leads to a new way of analyzing the origins of irreversibility in interacting systems, which we apply to the neural representation of visual inputs.

Evidence for the arrow of time arises because the probability of observing a trajectory and its time reverse are different, consistently. The proper information--theoretic measure of this difference is the Kullback--Leibler (KL) divergence \cite{Cover-01}. If we write trajectories schematically as $x(t)$, with $0 < t < T$, and the corresponding time reversed trajectories as $\tilde x(t)$, then the evidence for the arrow of time is
\begin{equation}
E \equiv D_{KL}\left( P[x(t)]\, ||\, P[\tilde x (t)] \right)=\sum_{x(t)} P[x(t)] \log \left( \frac{P[x(t)]}{P[\tilde x (t)]}\right).
\end{equation}
For large $T$ in steady--state systems, under mild assumptions, this evidence grows linearly with time, so it is natural to define a (global) rate
\begin{equation}
\dot I_{\rm global} \equiv \lim_{T\rightarrow\infty} \frac{E}{T}. 
\end{equation}
For a wide range of systems that can be described as making transitions coupled to a heat bath, this rate is equal to the rate $\dot S$ of entropy production that results from heat dissipation to the bath \cite{Seifert-01, Kawai-01}. To simplify the discussion, it is convenient to think of time advancing in discrete steps, and to consider observing just one transition, or two steps of the dynamics, at a time. This ``local arrow of time" or ``local irreversibility" can be written
\begin{equation}
\dot I = \sum_{x,x'} P(x\rightarrow x') \log\left[ \frac{P(x\rightarrow x')}{P(x'\rightarrow x)}\right],
\label{Idot_def}
\end{equation}
where
\begin{equation}
P(x\rightarrow x') = \text{Prob}(x_t = x,\, x_{t+1} = x').
\end{equation}
One can view this not as the Markovian approximation to the global irreversibility, but instead as the \textit{exact} evidence for the arrow of time contained in individual transitions. Equation (\ref{Idot_def}) makes very explicit that irreversibility is the breaking of detailed balance.

We are interested in systems where the state $x$ encompasses many interacting variables, $x \equiv \{x_{\rm i}\}$, with ${\rm i} = 1,\, 2,\, \cdots,\, N$. If we can measure dynamics with sufficient temporal resolution, then no two variables can change state at the same time.  Then instead of considering the full distribution $P(x\rightarrow x')$ we can focus on the $N$ individual distributions $P_{\rm i}(x_{\rm i}\rightarrow x_{\rm i}', \, x_{-{\rm i}})$, each of which describes a single variable transitioning from $x_{\rm i}$ to $x_{\rm i}'$ and the rest of the system remaining in the same state, denoted $x_{-{\rm i}}$. Such dynamics are referred to as multipartite, and exhibit a number of useful properties \cite{Horowitz-01, Horowitz-02, Wolpert-01}. Chief among these properties is the fact that the local irreversibility simplifies to a sum over the irreversibilities associated with the individual elements:
\begin{equation}
 \dot{I} = \sum_{{\rm i}=1}^N \dot{I}_{\rm i},
 \label{eq_I}
\end{equation}
where
\begin{equation}
\dot{I}_{\rm i} = \sum_{x_{-{\rm i}}} \sum_{x_{\rm i},x_{\rm i}'} P_{\rm i} (x_{\rm i}\rightarrow x_{\rm i}', \, x_{-{\rm i}}) \log \left[\frac{P_{\rm i} (x_{\rm i}\rightarrow x_{\rm i}', \, x_{-{\rm i}})}{P_{\rm i} (x_{\rm i}' \rightarrow x_{\rm i}, \, x_{-{\rm i}})}\right].
\label{eq_Si}
\end{equation}

If the different variables in the system are independent of one another, then the dynamics are fully defined by the marginal probabilities $P_{\rm i} (x_{\rm i}\rightarrow x_{\rm i}') = \sum_{x_{-{\rm i}}} P_{\rm i} (x_{\rm i}\rightarrow x_{\rm i}', \, x_{-{\rm i}})$, leading to an irreversibility
\begin{equation}
\dot{I}^{\text{ind}} = \sum_{{\rm i}=1}^N \dot{I}^{\text{ind}}_{\rm i},
\label{eq_Sind}
\end{equation}
where
\begin{equation}
\dot{I}_{\rm i}^{\rm ind} = \sum_{x_{\rm i} , x_{\rm i}'} P_{\rm i} (x_{\rm i}\rightarrow x_{\rm i}')\log\left[ \frac{P_{\rm i} (x_{\rm i}\rightarrow x_{\rm i}')}{P_{\rm i} (x_{\rm i}'\rightarrow x_{\rm i})}\right].
\end{equation}
Even if the variables are not independent we can always define this ``independent irreversibility''. The difference between this and the true irreversibility is the result of interactions,
\begin{equation}
\dot{I}^{\rm int} \equiv \dot{I} - \dot{I}^{\rm ind} = \sum_{{\rm i}=1}^N \dot{I}^{\text{int}}_{\rm i},
\end{equation}
where
\begin{align}
\dot{I}^{\text{int}}_{\rm i} &= \dot{I}_{\rm i}- \dot{I}^{\text{ind}}_{\rm i} \\
&= \sum_{x_{-{\rm i}}} \sum_{x_{\rm i},x_{\rm i}'} P_{\rm i}(x_{\rm i}\rightarrow x_{\rm i}', \, x_{-{\rm i}}) 
\log \left[\frac{P_{\rm i}(x_{-{\rm i}}\,| \, x_{\rm i} \rightarrow x_{\rm i}')}{P_{\rm i} (x_{-{\rm i}}\,| \, x_i' \rightarrow x_{\rm i})} \right] \label{eq_Sint} \\
&= \sum_{x_{\rm i},x_{\rm i}'} P_{\rm i} (x_{\rm i} \rightarrow x_{\rm i}' ) D_{KL}^{\rm i},
\end{align}
\begin{equation}
D_{KL}^{\rm i} = D_{KL} \left[P_{\rm i}(x_{-{\rm i}}\,| \, x_{\rm i} \rightarrow x_{\rm i}')\, || \, P_{\rm i}(x_{-{\rm i}}\,| \, x_{\rm i}' \rightarrow x_{\rm i})\right],
\label{eq_DKLi}
\end{equation}
and $P_{\rm i} (x_{-{\rm i}}\,| \, x_{\rm i} \rightarrow x_{\rm i}') = P_{\rm i}(x_{\rm i}\rightarrow x_{\rm i}', \, x_{-{\rm i}}) / P_{\rm i}(x_{\rm i}\rightarrow x_{\rm i}')$. Since each $D_{KL}^{\rm i} \geq 0$, this demonstrates that $\dot{I}^{\text{int}}_{\rm i} \ge 0$, so that interactions can only increase the irreversibility of a system. 

Equation (\ref{eq_Sint}) admits a simple interpretation: interactions contribute to the arrow of time if the observation of $x_{\rm i} \rightarrow x_{\rm i}'$ as opposed to $x_{\rm i}' \rightarrow x_{\rm i}$ points toward different states $x_{-{\rm i}}$ of the rest of the system. Thus, if $\rm i$'s forward- and reverse-time dynamics contain the same information about the rest of the system, then interactions do not contribute to $\rm i$'s local irreversibility ($\dot{I}^{\text{int}}_{\rm i} = 0$), and violations of detailed balance can only arise from independent dynamics ($\dot{I}_{\rm i} = \dot{I}^{\text{ind}}_{\rm i}$).

Together, Eqs.~(\ref{eq_Sind}-\ref{eq_DKLi}) establish our first main result: that the local arrow of time can be split into two non--negative components,
\begin{equation}
\label{eq_Sii}
\dot{I} = \dot{I}^{\text{ind}} + \dot{I}^{\text{int}},
\end{equation}
where $\dot{I}^{\text{ind}}$ reflects the local irreversibility of the individual elements and $\dot{I}^{\text{int}}$ reflects the local irreversibility due to interactions among the elements. Notice that for binary or Ising variables in steady state, we must have $P_{\rm i}(x_{\rm i}\rightarrow x_{\rm i}' ) = P_{\rm i}(x_{\rm i}' \rightarrow x_{\rm i})$ (such that $\dot{I}^{\text{ind}} = 0$), and so the local arrow of time necessarily arises from interactions ($\dot{I} = \dot{I}^{\text{int}}$). Additionally, we note that decomposition in Eq.~(\ref{eq_Sii}) requires multipartite dynamics; if multiple elements can change at once, then $\dot{I}^{\text{int}}$ is ill--defined (see Ref.~\cite{Lynn-12}).

When we say that interactions contribute to irreversibility, we have the intuition that these contributions can be further decomposed into interactions among pairs, triplets, etc.. Saying that we know only about interactions among pairs, for example, is equivalent to saying that we know all the marginal distributions
\begin{equation}
P_{\rm i} (x_{\rm i} \rightarrow x_{\rm i}',\, x_{\rm j}) = \sum_{x_{-\{{\rm i},{\rm j}\}}} P_{\rm i}(x_{\rm i} \rightarrow x_{\rm i}',\, x_{-{\rm i}}),
\label{eq_marg}
\end{equation}
where the sum runs over the states of all elements other than $\rm i$ and $\rm j$. We can then ask: what is the minimal irreversibility, or the weakest arrow of time, implied by these pairwise dynamics? To answer this question, we can search over all possible distributions $P_{\rm i} (x_{\rm i} \rightarrow x_{\rm i}', x_{-{\rm i}})$ that are consistent with the marginals in Eq.~(\ref{eq_marg}). Among these hypothetical systems, one will achieve a minimum of the local irreversibility in Eq.~(\ref{eq_I}), thus defining the minimum irreversibility consistent with the observed pairwise dynamics, denoted $\dot{I}^{(2)}$. This generalizes to higher orders, and since knowledge of $k^{\rm th}$--order dynamics includes all information about dynamics of lower orders $k' < k$, the result is a series of nested bounds,
\begin{equation}
\label{eq_hier}
0 \le \dot{I}^{(1)} \le \dot{I}^{(2)} \le \cdots \le \dot{I}^{(N-1)} \le \dot{I}^{(N)} = \dot{I},
\end{equation}
where we separately verify that $\dot{I}^{(1)} = \dot{I}^{\rm ind}$ \cite{Lynn-12}.

The hierarchy of bounds in Eq.~(\ref{eq_hier}) is analogous to the hierarchy of bounds on the entropy itself \cite{Schneidman-02}, but with inequalities reversed. When subjected to linear constraints on the underlying probability distributions---such as constraints on marginal distributions---entropy has a maximum while mutual information or KL divergence have a minimum \cite{Cover-01}. Colloquially, ``telling you more'' about the distribution increases the evidence for the local arrow of time. Note that computing each bound $\dot{I}^{(k)}$ requires finding a probability distribution that minimizes the irreversibility $\dot{I}$ subject to constraints on the $k^{\rm th}$--order dynamics \cite{Lynn-12}; because $\dot{I} = \dot{S}$ in thermodynamic contexts, this is equivalent to asking for dynamics that minimize entropy production. Minimizing entropy production is an idea that has been widely explored \cite{Schnakenberg-01, Wolpert-01, Skinner-01, Skinner-02, Still-01}, since the foundational work of Onsager and Prigogine \cite{Onsager-01, Prigogine-01}. Importantly we are not claiming that real systems minimize their entropy production, but rather are asking for hypothetical systems that have minimal entropy production consistent with a series of increasingly detailed constraints \cite{Skinner-01, Skinner-02}.

As a final interpretative step, it is natural to compare $\dot{I}^{(k)}$ with $\dot{I}^{(k-1)}$. If $\dot{I}^{(k)} = \dot{I}^{(k-1)}$, then the $k^{\text{th}}$--order dynamics are redundant in the sense that their irreversibility is entirely determined by lower--order correlations; by contrast, if $\dot{I}^{(k)} > \dot{I}^{(k-1)}$, then the $k^{\text{th}}$--order dynamics contain new information about the arrow of time. In this way, we can think about the difference between $\dot{I}^{(k)}$ and $\dot{I}^{(k-1)}$ as the contribution of interactions of order $k$ to the local arrow of time, $\dot{I}^{(k)}_{\text{int}} = \dot{I}^{(k)} - \dot{I}^{(k-1)} \ge 0$. This yields
\begin{align}
\label{eq_dec}
\nonumber \\[-30pt]
\dot{I} = \begin{array}{c} \\ \vspace{-4pt} \\ \underbrace{\begin{array}{c} \dot{I}^{(1)}_{\text{int}} \\ \vspace{-9pt} \\ \end{array}}_{\text{\normalsize $\dot{I}^{\text{ind}}$}} \end{array} + \begin{array}{c} \\ \vspace{-4pt} \\ \underbrace{\begin{array}{c} \dot{I}^{(2)}_{\text{int}} + \dot{I}^{(3)}_{\text{int}} + \cdots + \dot{I}^{(N)}_{\text{int}} \\ \vspace{-9pt} \\ \end{array}}_{\text{\normalsize $\dot{I}^{\text{int}}$}} \end{array} ,
\end{align}
which is our central result: the local arrow of time can be decomposed into non--negative contributions from individual elements in the system, interactions between pairs of elements, interactions among triplets, and so on.

\begin{figure}
\centering
\includegraphics[width = \columnwidth]{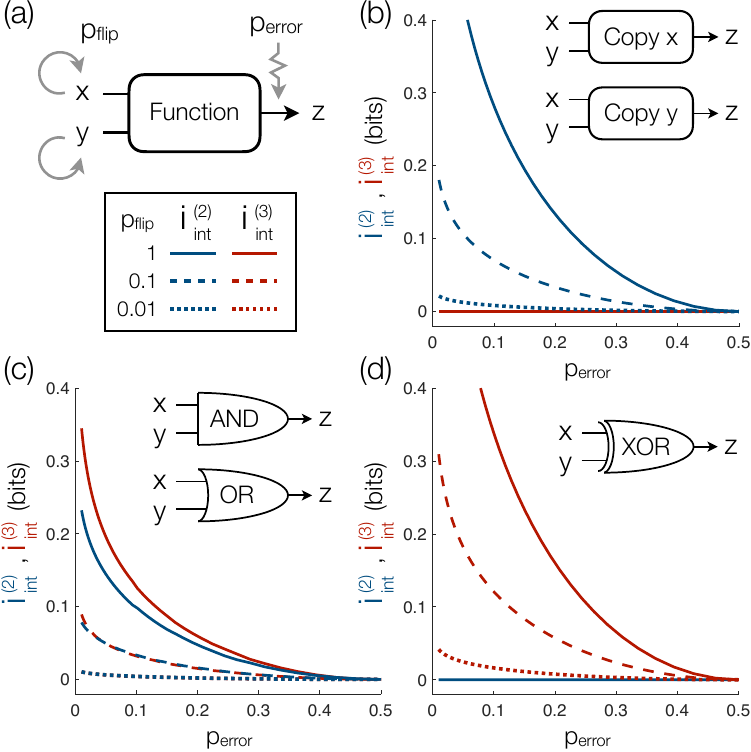} \\
\caption{Decomposing the irreversibility of logical functions. (a) System of three binary variables $x$, $y$, and $z$, where $z$ performs a noisy logical function on the inputs $x$ and $y$. At each point in time, one of the variables is updated at random. With probability $p_{\text{flip}}$, the inputs $x$ and $y$ change value, and with probability $p_{\text{error}}$, the output $z$ fails to perform the specified function. (b-d) Pairwise and triplet contributions to the local arrow of time  for different logical functions. Across all functions, irreversibility decreases with $p_{\text{error}}$ and increases with $p_{\text{flip}}$. (b) When $z$ copies either $x$ or $y$, the triplet irreversibility vanishes and all irreversibility arises from pairwise dynamics. (c) For AND and OR, irreversibility is driven by both pairwise and triplet dynamics. (d) For XOR, the pairwise irreversibility vanishes and all irreversibility arises from triplet dynamics. \label{fig_logical}}
\end{figure}

To illustrate this decomposition, we consider a minimal system of three binary variables $x, y, z$. At each moment in time, $z$ is a noisy logical function of $x$ and $y$, and from one time step to the next the variables $x$ and $y$ flip between their two states with probability $p_{\text{flip}}$, as in Fig.~\ref{fig_logical}(a). These dynamics are Markovian, so the local arrow of time is also the true global arrow of time. Since the variables are steady--state and binary, we have $\dot{I}^{\rm ind} = 0$, and since there are only three variables, the possible contributions are $\dot{I}_{\rm int}^{(2)}$ and $\dot{I}_{\rm int}^{(3)}$.

To begin, consider the simplest function, where $z$ copies either $x$ or $y$ while ignoring the other input [Fig.~\ref{fig_logical}(b)]. As $p_{\text{error}}$ increases (that is, as the accuracy of the function decreases), we find that the irreversibility $\dot{I}$ decreases, until at $p_{\text{error}} = 1/2$ (when the output $z$ completely decouples from the inputs $x$ and $y$) the system becomes reversible ($\dot{I} = 0$). Additionally, the irreversibility vanishes if the inputs $x$ and $y$ are static ($p_{\text{flip}} = 0$) and grows as the inputs become more dynamic ($p_{\text{flip}}$ increases). Notably, for all values of $p_{\text{flip}}$ and $p_{\text{error}}$, we find that $\dot{I}^{(3)}_{\text{int}} = 0$, and thus the irreversibility of the system arises entirely from pairwise dynamics, $\dot{I} = \dot{I}^{(2)}_{\text{int}}$.

For comparison, consider the AND, OR, and XOR functions [Figs.~\ref{fig_logical}(c) and (d)]. As before, the irreversibility increases as the functions become more accurate and as the inputs become more dynamic. In contrast to the copy functions, however, the irreversibilities of AND and OR (which are identical) arise in nearly equal amounts from pairwise and triplet interactions [Fig.~\ref{fig_logical}(c)]. Indeed, for both AND and OR, the output $z$ tends to increase with each of the inputs independently (yielding pairwise irreversibility), yet the full dynamics are not defined until all three variables are taken into account (yielding triplet irreversibility). The XOR function is the classic example of an irreducibly combinatorial interaction, so that the behavior of the system only becomes apparent once all three variables are observed simultaneously. As such, the pairwise dynamics are completely reversible, and all of the irreversibility is driven by triplet dynamics, so that $\dot{I} = \dot{I}^{(3)}_\text{int}$, as shown in Fig.~\ref{fig_logical}(d). Together, the results in Fig.~\ref{fig_logical} demonstrate how the local arrow of time can emerge from different orders of interactions among the many variables, even in relatively simple cases.

\begin{figure}
\centering
\includegraphics[width = \columnwidth]{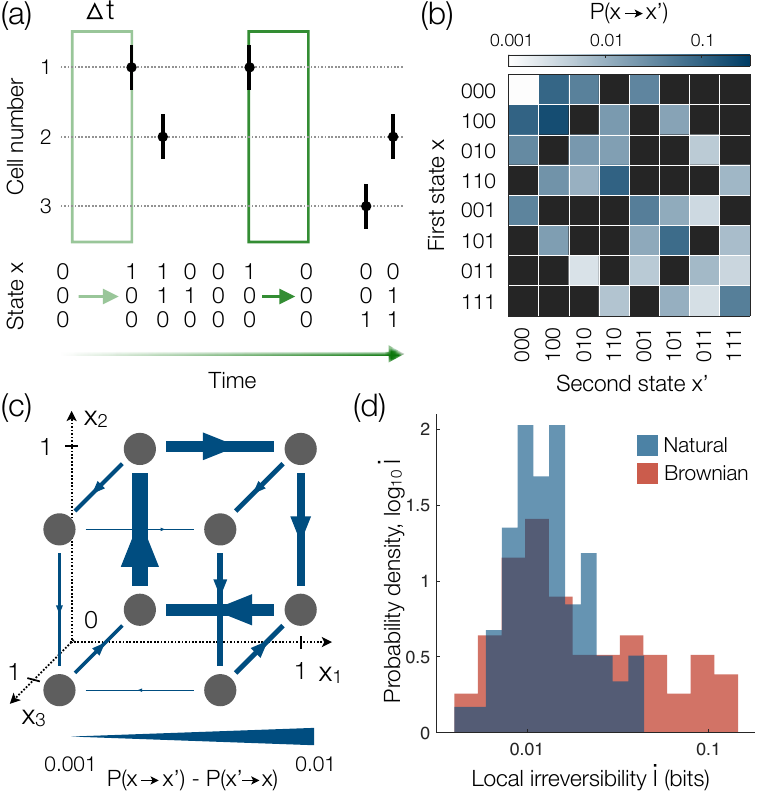} \\
\caption{Local irreversibility in groups of neurons. (a-c) A group of three neurons in the salamander retina responding to a natural movie. (a) Points mark the times of action potentials in each of three neurons. To define neural states, we slide a window of width $\Delta t = 20$ ms forward in time. When a cell generates an action potential inside the window we assign $x_{\rm i} = 1$, and when it is silent we assign $x_{\rm i} = 0$. Transitions occur when a spike enters the front or exits the back of the window, as shown. (b) The distribution $P(x\rightarrow x')$ over pairs of the eight possible states. Black entries indicate disallowed transitions. (c) Fluxes of probability $P(x\rightarrow x') - P(x'\rightarrow x)$ between neural states, demonstrating that this group of neurons breaks detailed balance. (d) Distributions of local irreversibilities $\dot{I}$ over 5-cell groups responding either to a natural movie (blue) or a Brownian stimulus (red), which is designed to be time-reversal invariant. Out of 100 random 5-cell groups, we only plot those ($\sim 2/3$) with significant local irreversibility. \label{fig_neuronal1}}
\end{figure}

Our perception of the arrow of time is driven by patterns of activity in networks of neurons. In particular, our experience of the visual world is constructed from the activity of retinal ganglion cells in the optic nerve, carrying information from eye to brain. Experimental developments have made it possible to record, simultaneously, from larger numbers of these cells, to the point that we can monitor all of the information that the brain receives about a small patch of the visual world \cite{Marre-01}. Here we focus on one such experiment, monitoring the activity of fifty--three neurons in the salamander retina as it responds to movies with different statistical structure \cite{Palmer-01}. Neurons have a naturally binary response, generating an action potential (spike; $x_{\rm i} =1$) or remaining silent ($x_{\rm i} =0$) in each small window of time ($\Delta t = 20\,{\rm ms}$), as in Fig.~\ref{fig_neuronal1}. As we slide the window forward in time, the time resolution of detecting action potentials is high enough that the dynamics are multipartite, so that a spike from only one neuron will enter the front of the window ($x_{\rm i} = 0\rightarrow 1$) or exit the back ($x_{\rm i} = 1\rightarrow 0$) at a time, as in Fig.~\ref{fig_neuronal1}(a). 

Figure \ref{fig_neuronal1}(b) shows the probabilities $P(x\rightarrow x')$ for a group of three neurons as they respond to a naturalistic movie, and Fig.~\ref{fig_neuronal1}(c) shows the corresponding fluxes of probability between states. Qualitatively we see that there are loops of flux, a hallmark feature of broken detailed balance in steady--state systems \cite{Battle-01, Gnesotto-01, Schnakenberg-01, Li-01}. In fact, we have checked that the neurons are in steady state \cite{Lynn-12}, such that the observed irreversibility cannot be attributed to non--stationarity. As with other information--theoretic quantities, estimating the local irreversibility from data is challenging, and prone to systematic errors due to finite data; we find that these can be controlled using the strategy of Ref.~\cite{Strong-01} if we restrict our attention to groups of no more than five cells. These and other technical details will be addressed in a longer paper \cite{Lynn-12}. When we choose groups of five cells at random from the fifty--three in the experimental population, more than two--thirds exhibit values of $\dot{I}$ that are significantly different from zero, with the distribution shown in Fig.~\ref{fig_neuronal1}(d).

Movies taken from natural settings, which provide the visual stimuli for the experiments analyzed here, obviously break time--reversal invariance. But one can construct movies for which the time--reverse is equally likely, for example a single bar moving along the trajectory generated by equilibrium Brownian motion of a damped harmonic oscillator, as was done (for different reasons) in the experiments of Ref.~\cite{Palmer-01}. Among the same groups of five neurons, we again find that roughly two--thirds exhibit nonzero local irreversibility, but perhaps surprisingly, these groups are more irreversible during the Brownian stimulus than the natural movie [Fig.~\ref{fig_neuronal1}(d)]. This difference in local irreversibility holds for all group sizes from $N=2$ to $N=5$. Together, these results demonstrate that retinal neurons break detailed balance in a way that does not simply reflect the irreversibility of the stimulus. In fact, the neuronal dynamics can define an arrow of time, even when the stimulus does not.

\begin{figure}
\centering
\includegraphics[width = \columnwidth]{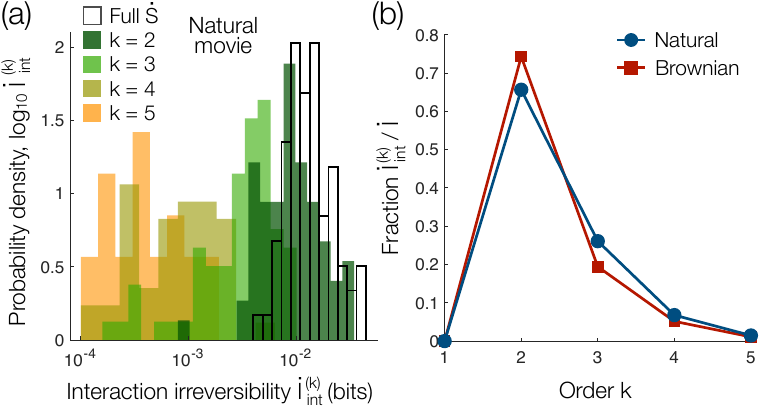} \\
\caption{Decomposing local irreversibility in neuronal activity. (a) Distributions of interaction irreversibilities $\dot{I}^{(k)}_{\text{int}}$ of different orders $k$ for random groups of five neurons responding to the natural movie.  (b) Interaction irreversibility $\dot{I}^{(k)}_{\text{int}}$, normalized by the full local irreversibility $\dot{I}$, as a function of the order $k$, averaged over the same 5-cell groups. \label{fig_neuronal2}}
\end{figure}

The decomposition in Eq.~(\ref{eq_dec}) gives us the opportunity to ask how the local arrow of time in neural responses is distributed across dynamics of different order. We recall that the interaction irreversibilities $\dot{I}^{(k)}_{\text{int}}$ fundamentally quantify the local arrow of time in $k^{\text{th}}$--order dynamics, and need not be driven by direct connections between neurons themselves. As noted above, the irreversibility that we observe is not the result of non--stationarity; more rigorously we find that in all the data we consider here, distributions really are in steady state within experimental error \cite{Lynn-12}. But for steady--state binary systems, the individual neurons cannot establish a local arrow of time ($\dot{I}^{\rm ind} = 0$); note that extended sequences of transitions from a single cell could generate irreversibility, but we focus here only on the local term. Thus, any irreversibility necessarily arises from statistical dependencies between two or more neurons at a time. For the same groups of $N=5$ cells in Fig.~\ref{fig_neuronal1}(d) responding to the natural movie, we find that pairwise dynamics account for much more of the local irreversibility than complex higher--order dynamics [Fig.~\ref{fig_neuronal2}(a)]. In fact, for both the natural and (time--reversal invariant) Brownian movies, pairwise dynamics contribute $66-74\%$ of the local irreversibility, more than any of the higher order terms [Fig.~\ref{fig_neuronal2}(b)].

To summarize, we have shown how evidence for the local arrow of time accumulates from the behavior of individual degrees of freedom and their interactions. Progressively higher--order dynamics each make a non--negative contribution, adding to the local irreversibility. As a practical matter, this decomposition allows us to (lower) bound the local irreversibility through measurements of low--order correlations in many--body systems, in much the same way that the maximum entropy method allows us to (upper) bound the entropy itself. We have focused here on the magnitude and decomposition of the local arrow of time, but it would be interesting to explore the hierarchy of minimally irreversible models that we construct along the way, especially as models for living systems. Perhaps these will be as successful in describing dynamics as the maximum entropy models have been in describing distributions of states at single moments in time \cite{Schneidman-01,Bialek-02,Meshulam-17,Lynn-04,Russ-21}. It would be interesting to understand the relationship of the minimally irreversible models to maximum entropy models for trajectories, sometimes called maximum caliber \cite{Presse-13,Cavagna-01}.

As a first step we have used our decomposition to analyze the responses of small groups of neurons in the retina as they encode complex visual inputs. It is not surprising that this initial neural representation of the visual world defines an arrow of time, although it is reassuring that this can be quantified, reliably. It is perhaps surprising that irreversibility is stronger in response to inputs that obey detailed balance, raising questions about how our internal perception of the arrow of time becomes aligned with the external world. Despite these large differences in the strength of the local arrow of time in response to different inputs, the way in which large-scale irreversibility is built out of fine-scale dynamics is constant, with the dominant role played by correlations among pairs of neurons. This relative simplicity holds out promise for simplified models of the neural dynamics, similar to pairwise maximum entropy models in the study of steady--state distributions. Generally, the emergence of irreversibility from pairwise dynamics opens the door for future investigations into whether, and how, the physical connections between neurons combine to produce a collective arrow of time. \\

\begin{acknowledgments}
We thank SE Palmer for helpful discussions and for guiding us through the data of Ref.~\cite{Palmer-01}. This work was supported in part by the National Science Foundation, through the Center for the Physics of Biological Function (PHY--1734030) and a Graduate Research Fellowship (CMH); by the National Institutes of Health through the BRAIN initiative (R01EB026943); by the James S McDonnell Foundation through a Postdoctoral Fellowship Award (CWL); by the Simons Foundation; and by a Sloan Research Fellowship (DJS).
\end{acknowledgments}

\textit{Citation diversity statement.}---Recent work in several fields of science \cite{Mitchell-01, Dion-01, Caplar-01, Dworkin-01, Bertolero-01}, and physics in particular \cite{Teich-01}, has identified citation bias negatively impacting women and other minorities. Here we sought to proactively consider choosing references that reflect the diversity of the field in thought, form of contribution, gender, and other factors. Excluding (including) self-citations to the current authors, our references contain $19\%$ ($25\%$) women lead authors and $31\%$ ($30\%$) women senior authors.

\bibliography{DetailedBalanceBib}

\end{document}